\documentclass[preprint,tightenlines,amsfonts,amsmath,amssymb,groupeaddress,showpacs,showkeys]{revtex4}

\usepackage{epsf} 
\usepackage{times}
\usepackage[latin1]{inputenc}
\usepackage{bm}
\usepackage{hyperref}  
\usepackage{latexsym}
\usepackage{graphicx}

\begin{document}

\title{Similar impact of topological and dynamic noise\\on complex patterns}
\author{Carsten \surname{Marr}} 
\email{marr@bio.tu-darmstadt.de}
\author{Marc-Thorsten \surname{H\"utt}} 
\affiliation{Bioinformatics Group, Department of Biology, Darmstadt
  University of Technology, 64287 Darmstadt, Germany}

\date{\today}

\begin{abstract}
  Shortcuts in a regular architecture affect the information transport
  through the system due to the severe decrease in average path
  length. A fundamental new perspective in terms of pattern formation
  is the destabilizing effect of topological perturbations by
  processing distant uncorrelated information, similarly to stochastic
  noise. We study the functional coincidence of rewiring and noisy
  communication on patterns of binary cellular automata.
\end{abstract}


\maketitle

Within the last two years cellular automata (CA) have re-emerged in
the focus of scientific attention, both with respect to fundamental
work \cite{matache05, sakai04} and with respect to applications, e.g.,
to biology \cite{bub05, sawai05, peak04}.  Such CA play an important
role as minimal models of complex pattern formation.  Pattern
complexity is usually discussed in terms of the four Wolfram classes
\cite{wolfram83} distinguishing fixed points, periodic, chaotic and
long-range complex behavior, respectively. Recently this approach to
pattern formation has been linked with concepts from graph theory by
studying the behavior of CA under variation of the underlying topology
\cite{marr05,moreira04}. Investigations elucidating the link between
dynamics and topology in complex networks are needed to ultimately
understand the evolutionary building principles of biological
networks.  The incorporation of noise or disorder in such systems is a
natural step towards realistic models \cite{moreira04,amaral04}.
Qualitatively speaking, there are two possible ways of introducing
disorder: On the level of signals and on the level of architecture.
The function of shortcuts in regular graphs with local clustering, as
first proposed in \cite{watts98}, has mostly been attributed to the
dramatic decrease of path lengths in these small-world networks. In
addition, however, topological perturbations can distribute global
information among the local neighborhoods. Recently, simple binary
dynamics in a noisy environment proved to be an excellent example for
efficient information processing in decentralized systems
\cite{moreira04}.

In this paper, we compare the two types of disorder in a model system
of complex pattern formation, namely a totalistic CA generalized to
arbitrary system topologies. We show that topological disorder, while
retaining determinism of the time development, produces qualitatively
the same effect on complex patterns as stochastic (external)
noise by inserting uncorrelated information into neighborhood-based
local patterns. We argue that this mechanism is responsible for the possibility to
pass from one Wolfram class to another purely by topological
modifications of the underlying graph that has been recently observed in \cite{marr05}. We show that this
relation between topological and dynamic noise remains, even when one
skips to a Monte Carlo update scheme.

\begin{figure}[t]
  \centering \epsfxsize=8cm \epsffile{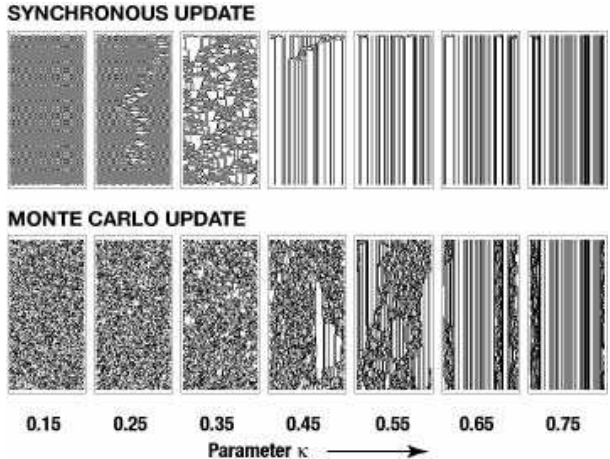} 
  \caption{Comparison of a synchronous (top) and a Monte Carlo (bottom)
    update scheme of the dynamics $\Omega(\kappa)$ on a regular graph
    with $d=10$ and $N=200$ for 400 time steps. A large $\kappa$ favors stationary behavior
    while small $\kappa$ values lead to an oscillatory and chaotic
    domain respectively. Complex Patterns emerge in both cases for
    intermediate values of the parameter $\kappa$. Here and in all
    following figures time runs in each spatiotemporal pattern
    downwards while the nodes are lined up horizontally according to
    their positions in the unperturbed regular network.}
  \label{updateschemes}
\end{figure}

We consider a simple binary dynamic $\Omega(\kappa)$ presented in
\cite{marr05} which is capable of generating complex behavior on a
standard CA topology, i.e.\ a regular one-dimensional graph with
periodic boundary conditions and $N$ nodes, where every node is
connected to its $d$ nearest neighbors. The state $x_i \in \{0,1\}$ of
a node $i$ is flipped or retained during the update process according
to the parameter $\kappa$ and the density $\rho_i$ of 1's in the
node's neighborhood, $\rho_i = \frac{1}{d_i} \sum_{j} A_{i\!j}
x_j(t)$, where $d_i$ is the degree of node $i$ and $A_{i\!j}$ is
the adjacency matrix of the underlying graph:
\begin{equation}
  \label{omega}
  \Omega(\kappa): \; x_i(t+1) = 
  \begin{cases}
    x_i(t) \,,    &  \rho_i \leq \kappa \\[0.2cm]
    1-x_i(t) \,,  &  \rho_i > \kappa \;.
    \end{cases}
\end{equation}  
The dynamic regimes for random initial conditions are intuitive:
Small $\kappa$ favor changing nodes while large $\kappa$ lead to
stationary behavior. In between these limiting regions, space time
patterns with both oscillatory and stationary elements occur and complex
dynamics can emerge.
Fig.~\ref{updateschemes} shows the $\kappa$ dependent domains for a
synchronous and a Monte Carlo update scheme on a regular graph. In the
first case, the simultaneous application of $\Omega(\kappa)$ to all
$N$ nodes in one time step leads to strictly oscillatory behavior around
small $\kappa$ and a narrow regime of complex patterns for $\kappa
\approx 0.35$. The Monte Carlo update of $\Omega(\kappa)$, where in
one time step $N$ randomly selected nodes are updated in random order,
leads to chaotic patterns for small $\kappa$ and a larger complex
region. In both cases, the variation of $\kappa$ results in a gradual
change of the dynamic complexity.

\begin{figure}[t]
  \centering  \epsfxsize=8cm \epsffile{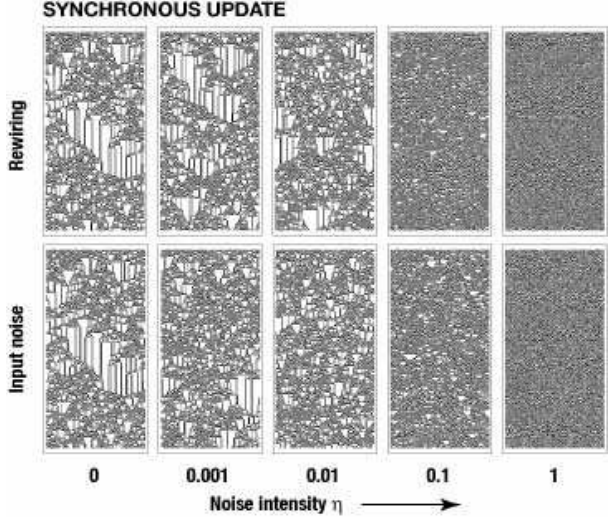}
  \caption{Spatiotemporal patterns for different degrees of
    topological and dynamic noise. Both sets of patterns are generated
    by a synchronous update of rule $\Omega(\kappa=0.35)$ on a
    directed graph with 200 nodes and $d=10$ for 400 time steps. They
    show a decrease of long-range correlations with increasing noise
    intensity $\eta$. For topological perturbations (top) the noise
    intensity $\eta$ specifies the fraction of rewired links.
    In the context of input noise (bottom), $\eta$ gives the
    probability for a neighbor being replaced by a random binary
    number during the update process.  Visually, both noise mechanisms
    have the same effect on the pattern formation capacity of the
    network dynamics.}
  \label{synchronPatterns}
\end{figure}
We systematically studied the effect of noise on the pattern formation
capacity of a synchronously updated network according to the rules
$\Omega(\kappa)$. In the noiseless case each $\Omega(\kappa)$ then
corresponds to a specific deterministic CA rule number. For disordered
graphs, we exploit the portability of $\Omega$ to arbitrary
architectures. In the following we choose a value of $\kappa$ giving
complex (Wolfram Class IV) patterns for the topologically and
dynamically unperturbed system. In Fig.~\ref{synchronPatterns} two
sets of spatiotemporal patterns are shown as typical examples for the
impact of noise on such CA dynamics. In the first case, the regular
graph topology is altered by rewiring the endpoints of randomly
selected links. This procedure is similar to the one described in
\cite{watts98}. Here, however, it is applied to directed graphs. 
In this context, the noise intensity $\eta$ specifies
the fraction of rewired connections. In the second case we surmise
that the communication between the elements is corrupted by noise: The
state $x_j$ of a linked element $j$ is substituted randomly with a
random binary number with probability $\eta$ during the update of
element $i$. We call this perturbation \textit{input noise} throughout
this paper, since it affects only the perception of the neighboring state
$x_j$ as seen by $i$ in Eq.~(\ref{omega}) and does not change the
state of $j$ itself.  The visual impression of
Fig.~\ref{synchronPatterns} suggests that input noise on the one hand
and rewiring the regular graph architecture on the other affects the
complexity of the system dynamics in a similar way.
\begin{figure}[t]
  \centering \epsfxsize=8cm \epsffile{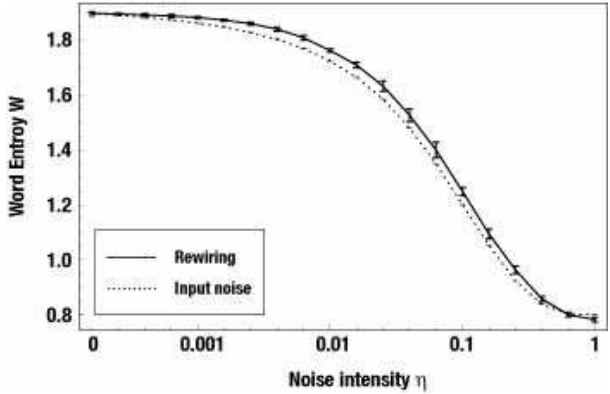}
  \caption{The word entropy $W$ as a function of the noise intensity
    $\eta$ for topological and dynamic noise on a synchronously
    updated network. We analyzed 10 runs of
    $\Omega(\kappa=0.35)$ for 500 nodes and 50000 time steps where a
    transient period of 1000 time steps has been dropped. As in
    Fig.~\ref{synchronPatterns}, the noise intensity $\eta$
    specifies the fraction of rewired links (solid line) and the
    probability for the occurrence of input noise during the update
    process (dotted line) respectively. For visual clarity, data
    points have been joined by lines.}
  \label{wordEntropy}
\end{figure}
We apply the word entropy $W$, introduced in \cite{marr05} and
tailored to the classification of CA into Wolfram classes, to quantify
the pattern formation capacity of the noisy network dynamics. The word
entropy $W_i$ relies on the relative number of words, i.e.\ blocks
of constant cell states, of different length in the time series $\{
x_i \}$ of a node $i$. The overall word entropy $W$ is the average
over all $N$ nodes of the graph,
\begin{equation}
  \label{word}
  W = \frac{1}{N} \sum_{i=1}^{N} W_i = 
  - \frac{1}{N} \sum_{i=1}^{N} \sum_{l=1}^{t} p_i(l) \log_2 p_i(l) \; ,
\end{equation}
where $p_i(l)$ is the probability for the occurrence of a word of
length $l$ in the time series of node $i$. For the $\kappa$-dependence
of $W$, see \cite{marr05}. In Fig.~\ref{wordEntropy},
we show the word entropy $W$ for systems perturbed by rewiring
and input noise, respectively. The pattern complexity decreases in both cases
similarly with the noise intensity $\eta$. Note that in both cases the
numerical value of the noise intensity gives the average number of
perturbations per link and time step.

\begin{figure}[t]
  \centering  \epsfxsize=8cm \epsffile{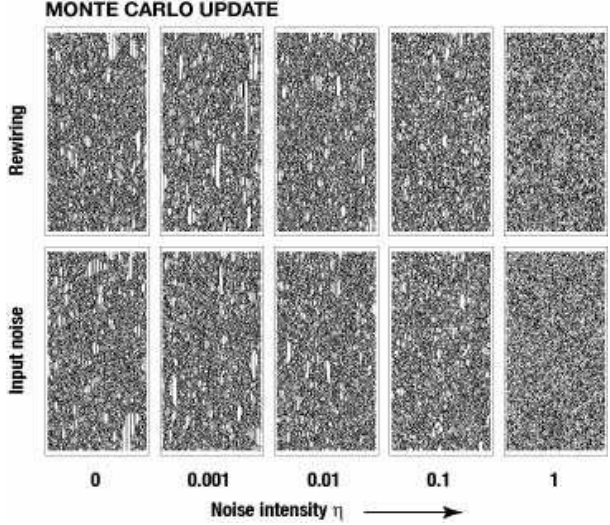}
  \caption{Spatiotemporal patterns of a randomly updated network
    for $\Omega(\kappa=0.4)$ on a directed graph with 200
    nodes and $d=10$. Spatial correlations vanish as $\eta$ is increased.}
  \label{mcPatterns}
\end{figure}
\begin{figure}[t]
  \centering  \epsfxsize=8cm \epsffile{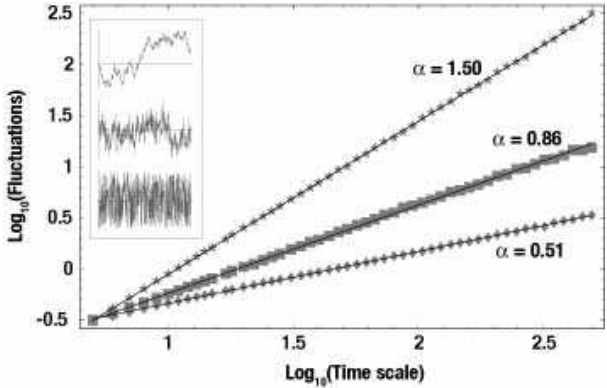}
  \caption{The detrended fluctuation analysis (DFA) quantifies the size of
    fluctuations in a signal on different time scales. The inset shows
    the time series of three signals analyzed in order to illustrate
    the different correlation strengths. A pure random, white noise
    signal (bottom) results in a slope $\alpha$ of about 0.5 in the
    log-log plot of fluctuations vs.\ time scale. An integrated white
    noise signal (top), called Brownian noise and representing a
    random walk leads to an $\alpha$ of 1.5. The signal in between
    stems from the density $\rho$ of a Monte Carlo updated automaton
    with $\kappa = 0.45$ shown in Fig.~\ref{updateschemes}. The
    $\alpha$ of 0.86 indicates the existence of long-range
    correlations, which can also be seen in the corresponding time
    series in between Brownian and white noise in the inset.}
  \label{dfaAnalyse}
\end{figure}

In a second investigation we studied the same phenomenon under a Monte
Carlo update scheme. This random time evolution is argued to be more
realistic in terms of information processing in natural systems, as a
simultaneous update may induce stable patterns as artifacts.
As $\eta$ is increased, coherent regions in the patterns disappear and
chaotic behavior prevails (Fig.~\ref{mcPatterns}). Due to the change
in the update scheme, a proper separation of the dynamic regimes with
the word entropy fails: Short $l$-words of different lengths in the
chaotic patterns result in similarly high $W$ values as in the complex
case with long-range correlations. In order to definitely classify the
pattern evolution with a single observable one could use the entropy
measure proposed by Wolfram in \cite{wolfram83}, where all possible
words, including non-constant $l$-words, are accounted for. Instead,
we will apply the (statistically less demanding) detrended fluctuation
analysis (DFA) to reveal the complexity of the binary network
dynamics, as applied in \cite{amaral04}.

The DFA characterizes the time correlations of a signal with a single
value, the scaling exponent $\alpha$. A random signal with no
long-range time correlations leads to $\alpha = 0.5$, while
Brownian noise exhibits an exponent of 1.5. Many physiologic systems
have been shown to generate signals with long-range time correlations
\cite{goldberger02} resulting in an intermediate scaling exponent of
$\alpha \approx 1$. Fig.~\ref{dfaAnalyse} shows three signals with
different time correlations and the result of an applied DFA.  The
known relationship to the scaling exponent $\alpha_{ps}$ of the
power spectrum of fluctuations, $\alpha_{ps} = 2 \alpha - 1$, allows a
comprehensive discussion of parallels between complex dynamics with
$\alpha = 1$ and natural phenomena with 1/$f$ fluctuations
\cite{bak87}.  Detailed descriptions of the DFA algorithm can be found
in \cite{peng94}.
\begin{figure}[t]
  \centering  \epsfxsize=8cm \epsffile{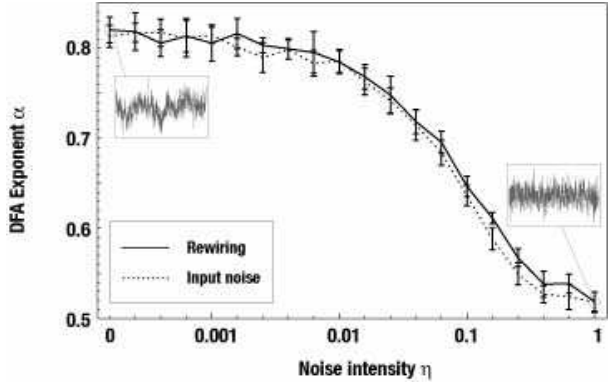}
  \caption{Detrended fluctuation analyses (DFA) of a Monte Carlo updated dynamics
    $\Omega(\kappa=0.4)$ against the noise intensity $\eta$. Both
    noise mechanisms have similar effects on the complexity of the
    time series of the density $\rho$ of the system. Again, we used a network
    consisting of 500 nodes with $d=10$ and analyzed the time series
    of 50000 time steps discarding a transient period.  The error bars
    indicate the deviations of 10 runs, where the random initial
    conditions and the graph randomizations have been varied. The time
    series in the insets show the correlation character of the
    analyzed signals.}
  \label{alpha}
\end{figure}
We use the DFA to measure time correlations in the overall density of
a system $\rho(t) = \frac{1}{N} \sum_{i=1}^N x_i(t)$ and thereby
quantify the complexity of the system's dynamics.  Again we find that
both, perturbations of the regular topology and dynamic noise affect
the system in a intriguingly similar manner (Fig.~\ref{alpha}).  In
the extremal case of $\eta=1$ (which corresponds to completely
disintegrated neighborhoods and the permanent substitution of
neighboring states with random binary numbers respectively) both
systems exhibit a completely random, white noise
signal, characterized by an exponent $\alpha \approx 0.5$.

\begin{figure}[t]
  \centering  \epsfxsize=7cm \epsffile{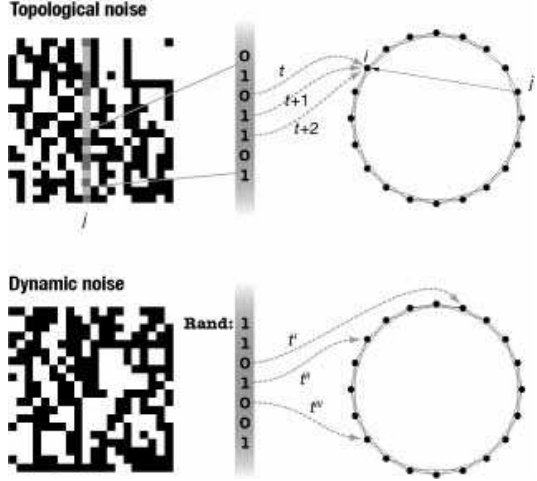}
  \caption{Qualitative visualization of the two noise mechanisms for a
    $N=50$, $d=4$ graph and a noise intensity of $\eta=1/(dN)$. The
    schematic diagrams on the left show the result of a network
    dynamic, a spatiotemporal pattern of zeroes and ones where time
    runs downward and the 50 nodes are lined up horizontally according
    to their position in the unperturbed chain. In the upper picture,
    a simple directed link has been rewired and accordingly, the
    states of a distant node $j$ affect the update of node $i$ through
    this shortcut in every time step $t, t+1, t+2, \dots$. The
    perturbing binary elements can be lined up as a noise vector,
    consisting of the time series of element $j$. In contrast, for
    dynamic perturbations this noise vector is made up of random
    binary numbers. These affect the update process of randomly chosen
    states at random times $t', t'', t''', \dots$, substituting the
    state of a neighboring node with probability $eta$.}
  \label{noises}
\end{figure}

How can the similarity of two such different noise mechanisms be
understood?  Complex dynamics given, shortcuts insert information from
distant regions of the network into local neighborhoods, which
is---due to its lack of correlation to the neighborhood at hand and
the overall dynamics of the network's elements---similar to random
dynamic perturbations. Fig.~\ref{noises} depicts the two discussed
noise mechanisms schematically. One prerequisite for the similar
impact of both noise mechanisms is therefore clearly a globally
running complex or chaotic dynamics, which ensures an adequately
random and uncorrelated signal from the linked distant node.
Apparently deviations in the space and time correlations of the noise
events are leveled by the internal signal processing of the network.
Finally, we want to note that other kinds of dynamical noise (like
white noise affecting directly the state $x_i$ of the system) and
other forms of topological noise (e.g.  the addition of links into the
architecture) lead to similar results as the mechanisms discussed
above.

The effect described in this paper may apply to a large class of
network dynamics phenomena. The results in \cite{graham03} for example,
where the forest-fire-model is discussed on a small-world topology
can be interpreted in this way. There, a self-organized critical state
of the system is achieved without the introduction of dynamic noise
(namely the occurrence of lightning strikes) by rewiring a small
fraction of links. On the other hand, topological and dynamic noise
can also play different functional roles in network dynamics, as
reported in \cite{moreira04} for the case of a simple density
classifier task. Indeed, the globalizing effect of shortcuts and the
random perturbations of input noise provide two very different
resources for an effective processing of the majority rule. However, complex dynamics can be regarded as a functional state of a
complex network itself, as discussed recently in \cite{zumdieck04}.
As we have shown, the systematic modification of such dynamics can be
equally achieved by dynamic noise and topological disorder. We expect
that this functional alternative in regulating dynamic features can be
found to be exploited in natural systems, as soon as one looks at
phenomena from the perspective lined out here.  

Summarizing, we find that in graphs with clustered neighborhood
structures, links between distant regions of the
network can have the same effect as stochastic perturbations of the
dynamic signals themselves, if the signal conferred by these shortcuts displays an
appropriate degree of chaoticity.


\begin{thebibliography}{16}
\expandafter\ifx\csname natexlab\endcsname\relax\def\natexlab#1{#1}\fi
\expandafter\ifx\csname bibnamefont\endcsname\relax
  \def\bibnamefont#1{#1}\fi
\expandafter\ifx\csname bibfnamefont\endcsname\relax
  \def\bibfnamefont#1{#1}\fi
\expandafter\ifx\csname citenamefont\endcsname\relax
  \def\citenamefont#1{#1}\fi
\expandafter\ifx\csname url\endcsname\relax
  \def\url#1{\texttt{#1}}\fi
\expandafter\ifx\csname urlprefix\endcsname\relax\def\urlprefix{URL }\fi
\providecommand{\bibinfo}[2]{#2}
\providecommand{\eprint}[2][]{\url{#2}}

\bibitem[{\citenamefont{Matache and Heidel}(2005)}]{matache05}
  \bibinfo{author}{\bibfnamefont{M.~T.} \bibnamefont{Matache}}
  \bibnamefont{and}
  \bibinfo{author}{\bibfnamefont{J.}~\bibnamefont{Heidel}},
  \bibinfo{journal}{Phys.~Rev.~E} \textbf{\bibinfo{volume}{71}},
  \bibinfo{pages}{026232} (\bibinfo{year}{2005}).


\bibitem[{\citenamefont{Sakai et~al.}(2004)\citenamefont{Sakai, Kanno, and
  Saito}}]{sakai04}
\bibinfo{author}{\bibfnamefont{S.}~\bibnamefont{Sakai}},
  \bibinfo{author}{\bibfnamefont{M.}~\bibnamefont{Kanno}}, \bibnamefont{and}
  \bibinfo{author}{\bibfnamefont{Y.}~\bibnamefont{Saito}},
  \bibinfo{journal}{Phys.~Rev.~E} \textbf{\bibinfo{volume}{69}},
  \bibinfo{pages}{066117} (\bibinfo{year}{2004}).

\bibitem[{\citenamefont{Bub et~al.}(2005)\citenamefont{Bub, Shrier, and
  Glass}}]{bub05}
\bibinfo{author}{\bibfnamefont{G.}~\bibnamefont{Bub}},
  \bibinfo{author}{\bibfnamefont{A.}~\bibnamefont{Shrier}}, \bibnamefont{and}
  \bibinfo{author}{\bibfnamefont{L.}~\bibnamefont{Glass}},
  \bibinfo{journal}{Phys. Rev. Lett.} \textbf{\bibinfo{volume}{94}},
  \bibinfo{pages}{028105}  (\bibinfo{year}{2005}).


\bibitem[{\citenamefont{Peak et~al.}(2004)\citenamefont{Peak, West, Messinger,
  and Mott}}]{peak04}
\bibinfo{author}{\bibfnamefont{D.}~\bibnamefont{Peak}},
  \bibinfo{author}{\bibfnamefont{J.~D.} \bibnamefont{West}},
  \bibinfo{author}{\bibfnamefont{S.~M.} \bibnamefont{Messinger}},
  \bibnamefont{and} \bibinfo{author}{\bibfnamefont{K.~A.} \bibnamefont{Mott}},
  \bibinfo{journal}{Proc.~Natl.~Acad.~Sci.~U.S.A.} \textbf{\bibinfo{volume}{101}}, \bibinfo{pages}{918}
  (\bibinfo{year}{2004}).

\bibitem[{\citenamefont{Sawai et~al.}(2005)\citenamefont{Sawai, Thomason, and
  Cox}}]{sawai05}
\bibinfo{author}{\bibfnamefont{S.}~\bibnamefont{Sawai}},
  \bibinfo{author}{\bibfnamefont{P.~A.} \bibnamefont{Thomason}},
  \bibnamefont{and} \bibinfo{author}{\bibfnamefont{E.~C.} \bibnamefont{Cox}},
  \bibinfo{journal}{Nature} \textbf{\bibinfo{volume}{433}},
  \bibinfo{pages}{323} (\bibinfo{year}{2005}).

\bibitem[{\citenamefont{Wolfram}(1983)}]{wolfram83}
\bibinfo{author}{\bibfnamefont{S.}~\bibnamefont{Wolfram}},
  \bibinfo{journal}{Rev. Mod. Phys.} \textbf{\bibinfo{volume}{55}},
  \bibinfo{pages}{601} (\bibinfo{year}{1983}).

\bibitem[{\citenamefont{Marr and Hütt}(2005)}]{marr05}
\bibinfo{author}{\bibfnamefont{C.}~\bibnamefont{Marr}} \bibnamefont{and}
  \bibinfo{author}{\bibfnamefont{M.-T.} \bibnamefont{Hütt}},
  \bibinfo{journal}{Physica A} \textbf{\bibinfo{volume}{354}},
  \bibinfo{pages}{641} (\bibinfo{year}{2005}).

\bibitem[{\citenamefont{Moreira et~al.}(2004)\citenamefont{Moreira,
      Mathur, Diermeier and Amaral}}]{moreira04}
  \bibinfo{author}{\bibfnamefont{A.~A.} \bibnamefont{Moreira}},
  \bibinfo{author}{\bibfnamefont{A.}~\bibnamefont{Mathur}},
  \bibinfo{author}{\bibfnamefont{D.}~\bibnamefont{Diermeier}},
  \bibnamefont{and} \bibinfo{author}{\bibfnamefont{L.~A.~N.}
    \bibnamefont{Amaral}}, \bibinfo{journal}{Proc.~Natl.~Acad.~Sci.~U.S.A.}
  \textbf{\bibinfo{volume}{101}}, \bibinfo{pages}{12085}
  (\bibinfo{year}{2004}).

\bibitem[{\citenamefont{Amaral et~al.}(2004)\citenamefont{Amaral, Díaz-Guilera,
  Moreira, Goldberger, and Lipsitz}}]{amaral04}
\bibinfo{author}{\bibfnamefont{L.~A.~N.} \bibnamefont{Amaral}},
  \bibinfo{author}{\bibfnamefont{A.}~\bibnamefont{Díaz-Guilera}},
  \bibinfo{author}{\bibfnamefont{A.~A.} \bibnamefont{Moreira}},
  \bibinfo{author}{\bibfnamefont{A.~L.} \bibnamefont{Goldberger}},
  \bibnamefont{and} \bibinfo{author}{\bibfnamefont{L.~A.}
  \bibnamefont{Lipsitz}}, \bibinfo{journal}{Proc.~Natl.~Acad.~Sci.~U.S.A.}
  \textbf{\bibinfo{volume}{101}}, \bibinfo{pages}{15551}
  (\bibinfo{year}{2004}).

\bibitem[{\citenamefont{Watts and Strogatz}(1998)}]{watts98}
\bibinfo{author}{\bibfnamefont{D.~J.} \bibnamefont{Watts}} \bibnamefont{and}
  \bibinfo{author}{\bibfnamefont{S.~H.} \bibnamefont{Strogatz}},
  \bibinfo{journal}{Nature} \textbf{\bibinfo{volume}{393}},
  \bibinfo{pages}{440} (\bibinfo{year}{1998}).
  
\bibitem[{\citenamefont{Goldberger
      et~al.}(2002)\citenamefont{Goldberger, Amaral, Hausdorff, Ivanov, Peng, and
      Stanley}}]{goldberger02} 
  \bibinfo{author}{\bibfnamefont{A.~L.} \bibnamefont{Goldberger}},
  \bibinfo{author}{\bibfnamefont{L.~A.~N.} \bibnamefont{Amaral}},
  \bibinfo{author}{\bibfnamefont{J.~M.} \bibnamefont{Hausdorff}},
  \bibinfo{author}{\bibfnamefont{P.~C.} \bibnamefont{Ivanov}}, 
  \bibinfo{author}{\bibfnamefont{C.-K.} \bibnamefont{Peng}}, \bibnamefont{and}
  \bibinfo{author}{\bibfnamefont{H.~E.} \bibnamefont{Stanley}},
  \bibinfo{journal}{Proc.~Natl.~Acad.~Sci.~U.S.A.} \textbf{\bibinfo{volume}{99}},
  \bibinfo{pages}{2466} (\bibinfo{year}{2002}).

\bibitem[{\citenamefont{Bak et~al.}(1987)\citenamefont{Bak, Tang, and
  Wiesenfeld}}]{bak87}
\bibinfo{author}{\bibfnamefont{P.}~\bibnamefont{Bak}},
  \bibinfo{author}{\bibfnamefont{C.}~\bibnamefont{Tang}}, \bibnamefont{and}
  \bibinfo{author}{\bibfnamefont{K.}~\bibnamefont{Wiesenfeld}},
  \bibinfo{journal}{Phys.~Rev.~Letters} \textbf{\bibinfo{volume}{59}},
  \bibinfo{pages}{381} (\bibinfo{year}{1987}).

\bibitem[{\citenamefont{Peng et~al.}(1994)\citenamefont{Peng, Buldyrev, Havlin,
  Simons, Stanley, and Goldberger}}]{peng94}
\bibinfo{author}{\bibfnamefont{C.-K.} \bibnamefont{Peng}},
  \bibinfo{author}{\bibfnamefont{S.~V.} \bibnamefont{Buldyrev}},
  \bibinfo{author}{\bibfnamefont{S.}~\bibnamefont{Havlin}},
  \bibinfo{author}{\bibfnamefont{M.}~\bibnamefont{Simons}},
  \bibinfo{author}{\bibfnamefont{H.~E.} \bibnamefont{Stanley}},
  \bibnamefont{and} \bibinfo{author}{\bibfnamefont{A.~L.}
  \bibnamefont{Goldberger}}, \bibinfo{journal}{Phys.~Rev.~E}
  \textbf{\bibinfo{volume}{49}}, \bibinfo{pages}{1685} (\bibinfo{year}{1994}).


\bibitem[{\citenamefont{Graham and Matthai}(2003)}]{graham03}
\bibinfo{author}{\bibfnamefont{I.}~\bibnamefont{Graham}} \bibnamefont{and}
  \bibinfo{author}{\bibfnamefont{C.~C.} \bibnamefont{Matthai}},
  \bibinfo{journal}{Phys.~Rev.~E} \textbf{\bibinfo{volume}{68}},
  \bibinfo{pages}{036109} (\bibinfo{year}{2003}).

\bibitem[{\citenamefont{Zumdieck et~al.}(2004)\citenamefont{Zumdieck, Timme,
  Geisel, and Wolf}}]{zumdieck04}
\bibinfo{author}{\bibfnamefont{A.}~\bibnamefont{Zumdieck}},
  \bibinfo{author}{\bibfnamefont{M.}~\bibnamefont{Timme}},
  \bibinfo{author}{\bibfnamefont{T.}~\bibnamefont{Geisel}}, \bibnamefont{and}
  \bibinfo{author}{\bibfnamefont{F.}~\bibnamefont{Wolf}},
  \bibinfo{journal}{Phys.~Rev.~Lett.} \textbf{\bibinfo{volume}{93}},
  \bibinfo{pages}{244103} (\bibinfo{year}{2004}),


\end{thebibliography}
\end{document}